\documentclass[pra,aps,twocolumn,superscriptaddress,tightenlines, longbibliography]{revtex4-2}
\usepackage{graphicx}
\usepackage{bm}
\usepackage{braket}
\usepackage{pifont}
\usepackage{amsmath}
\usepackage{amsfonts}
\usepackage{amssymb}    
\usepackage[english]{babel}
\usepackage{color,xcolor}
\usepackage[caption=false]{subfig}
\usepackage{tensor}
\usepackage{natbib}
\usepackage{dsfont}
\usepackage{hyperref}
\hypersetup{
	colorlinks=true,
	linkcolor=blue,
	filecolor=black,  
	citecolor=blue, 
	urlcolor=blue,
	pdftitle={Entanglement generation between distant spins via quasilocal reservoir engineering},
	bookmarks=true,
}
\newcommand{\cjosephine}[1]{{\color{black} {#1}}}

\graphicspath{{./../Figures/}}
\usepackage{cleveref}
\begin{document}
\title{Entanglement generation between distant spins via quasilocal reservoir engineering}
\author{Josephine Dias} 
\email{josephine.dias@oist.jp}
\affiliation{Okinawa Institute of Science and Technology Graduate University, Onna-son$,$ Okinawa 904-0495$,$ Japan}
\author{Christopher W. Wächtler} 
\affiliation{Institut für Theoretische Physik, Technische Universität Berlin, Hardenbergstr. 36, 10623 Berlin, Germany}
\affiliation{Department of Physics, University of California, Berkeley, California 94720, USA}
\author{Kae Nemoto}
\affiliation{Okinawa Institute of Science and Technology Graduate University, Onna-son$,$ Okinawa 904-0495$,$ Japan}
\affiliation{National Institute of Informatics, 2-1-2 Hitotsubashi, Chiyoda, Tokyo 101-0003, Japan.}
\author{William J. Munro}
\affiliation{Okinawa Institute of Science and Technology Graduate University, Onna-son$,$ Okinawa 904-0495$,$ Japan}
\affiliation{NTT Basic Research Laboratories $\&$ NTT Research Center for Theoretical Quantum Physics, NTT Corporation,  3-1 Morinosato-Wakamiya,  Atsugi,  Kanagawa 243-0198,  Japan.}
\affiliation{National Institute of Informatics, 2-1-2 Hitotsubashi, Chiyoda, Tokyo 101-0003, Japan.}
\date{\today}
\begin{abstract}
The generation and preservation of entanglement is a central goal in quantum technology. Traditionally, dissipation in quantum systems is thought to be detrimental to entanglement, however dissipation can also be utilised as a means of generating entanglement between quantum spins that are not directly interacting. In particular entanglement can be generated between two qubits, or multi qubit systems via a collective coupling to a reservoir. In this work, we explore multiple spin domains pairwise coupled to different reservoirs and show that entanglement can be generated between spins which are not coupled to each other, or even coupled to the same reservoir. 
\end{abstract}
\maketitle

\section{Introduction}
Quantum entanglement enables quantum computation and communication that surpass their classical counterparts in various tasks \cite{vidal2003efficient,ekert1998quantum, gottesman1999demonstrating,raussendorf2001one, gisin2007quantum, jozsa2003role}. Therefore, the generation and preservation of entanglement represents a significant goal within the field of quantum technology. Typically, dissipative effects are known to be detrimental to quantum entanglement \cite{zyczkowski2001dynamics}, and significant efforts are made into isolating the quantum system from its environment. However, the environment can also be used as a means of generating entanglement between different quantum systems \cite{rajagopal2001decoherence,hormeyll2009environment, braun2002creation,kim2002entanglement, schneider2002entanglement}.

With recent advancements in hybrid quantum systems, we can engineer both the system \textit{and} the environment to perform the task we are interested in \cite{verstraete2009quantum,keck2018persistent,damanet2019controlling,yanay2018reservoir}. Such hybrid systems can involve many different elements, from solid state to atomic, molecular and optical \cite{kurizki2015quantum, wallquist2009hybrid, xiang2013hybrid}. Hybrid quantum systems naturally provide a rich environment to study complex phenomena in open quantum systems \cite{BreuerPetruccioneBook2002}. Our focus in this work is the behavoir of many spin ensembles collectively coupled to bosonic reservoirs and the entanglement that can be generated through this system. The collective coupling of multiple spins to the same bath has been shown as a mechanism to transfer excitations between spins \cite{hama2018relaxation,hama2018negative, stegmann2020relaxation} and also as a means to  charge quantum batteries \cite{quach2020using}.  Furthermore mutliple spin domains pairwise coupled to mulitple reservoirs can transfer excitations to and from domains which are not directly coupled \cite{dias2021reservoir}. 

Open quantum systems form a complex and dynamic backdrop to study the generation and dynamics of entanglement (see Ref.~\cite{aolita2015open} for a review). It is known that entanglement can be generated between two qubits via the interaction of the qubits with the same reservoir \cite{hormeyll2009environment, braun2002creation,kim2002entanglement, oh2006entanglement, contreraspulido2008entanglement, maniscalco2008protecting,francica2009resonant,lin2013dissipative,shankar2013autonomously} or the interaction of two oscillators with the same reservoir \cite{benatti2006entangling}.    A collective coupling to a common bath can also be used to generate entanglement between multiple spins \cite{schneider2002entanglement} and spin domains \cite{hama2018negative}, even if the reservoir is in a thermal state \cite{braun2002creation, kim2002entanglement}. For instance, a system of two spin domains collectively coupled to the same bosonic reservoir was explored and its steady state was shown to have entanglement between spin-domains \cite{hama2018negative}.

In order to illustrate the mechanism by which initially separable systems can become entangled via a collective coupling to a reservoir, consider an elementary example of two spins $A$ and $B$ collectively coupled to the same \cjosephine{zero-temperature reservoir} and initialised in the separable state  $\ket{\psi}_{AB}= \ket{\uparrow}_A\otimes \ket{\downarrow}_B $. This state can also be written as $\ket{\psi}_{AB}=\frac{1}{\sqrt{2}}\left( \ket{\Psi^+}_{AB}+\ket{\Psi^-}_{AB}\right) $ which is an equal superposition of the symmetric triplet state $\ket{\Psi^+}_{AB}$ and antisymmetric singlet state $\ket{\Psi^-}_{AB}$. \cjosephine{Under the action of the collective dissipator $J_A^-+J_B^-$}, the triplet state decays to the ground state $\ket{\downarrow}_A\otimes \ket{\downarrow}_B$, however the singlet state cannot decay. Therefore, the steady state for this system is \cite{hormeyll2009environment}
\begin{equation}
\rho_{AB}=\frac{1}{{2}}\left( \ket{\downarrow}_A\ket{\downarrow}_B  \bra{\downarrow}_A\bra{\downarrow}_B +   \ket{\Psi^-}_{AB}  \bra{\Psi^-}_{AB}\right) 
\label{eq:0}
\end{equation}
and this is an entangled state between spins $A$ and $B$. If we were to quantify the amount of entanglement in this state, we could use the entanglement of formation \cite{bennett1996purification,wootters1998entanglement, hill1997entanglement} (see Sec.~\ref{sec:dynamics}) which for this two-qubit state is $E_F\approx0.354$, that is $0.354$ ebits.  We emphasise here that the spins were not initially entangled, nor did they evolve according to 
a Hamiltonian coupling individual spins. The entanglement present in the steady state of \eqref{eq:0} is solely due to the relaxation of the spins via their collective reservoir coupling. 

\begin{figure}
	\centering
	\includegraphics[width=0.83\linewidth]{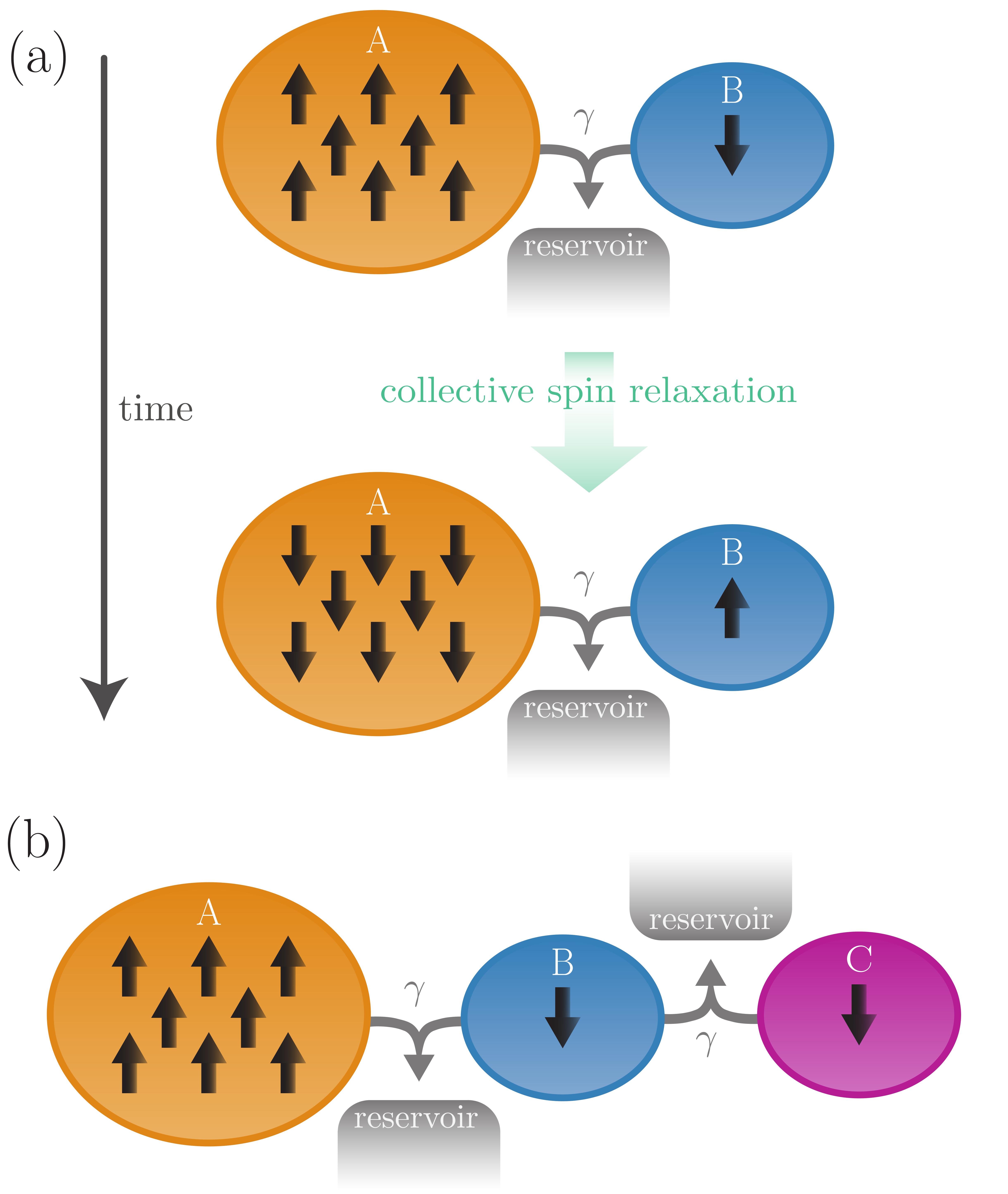}
\refstepcounter{subfigure}
	\caption{Collective spin relaxation via a collective coupling to a bosonic reservoir can be used to transfer excitations between spins. (a) The double domain system of Ref.~\cite{hama2018negative,hama2018relaxation} (b) Three spin domains pairwise coupled to two reservoirs. }
	\label{fig:1}
\end{figure}

To go beyond this simple example of two spins, consider the system depicted in Fig.~\ref{fig:1} (a) (which is the system considered in Refs.~\cite{hama2018relaxation, hama2018negative}). Here we have a system of $N_A$ spins in domain $A$, and a single spin in domain $B$. With this system initialised in the state of 
\begin{equation}
\ket{\psi}_{AB} = \ket{\uparrow\uparrow\cdots\uparrow}_A\otimes \ket{\downarrow}_B
\end{equation}
after collective spin relaxation (for a zero-temperature bath), the steady state will be
\begin{equation}
\ket{\psi}_{AB} = \ket{\downarrow\downarrow\cdots\downarrow}_A\otimes \ket{\uparrow}_B
\label{eq:s2}
\end{equation}
in the limit of $N_A\to \infty$. That is, through the collective coupling of the spins to the bosonic reservoir, the steady state of domain $B$ is excited even though it initially started in its ground state \cite{hama2018negative, hama2018relaxation}. As can be seen from \eqref{eq:s2}, the steady state in the limit of $N_A\to \infty$ is separable. However, for finite $N_A$, collective spin relaxation produces entanglement in the steady state between domains $A$ and $B$. In fact, the logarithmic negativity \cite{vidal2002computable, plenio2005logarithmic} for the steady state peaks for a spin population of $N_A=5$ and has a value of $E_N \approx 0.55$ \cite{hama2018negative, munro2021collective}. 

To illustrate how we can extend the model of Fig.~\ref{fig:1}(a) in Ref.~\cite{hama2018negative,hama2018relaxation}, consider the system depicted in Fig.~\ref{fig:1}(b):  In addition to the two spin domains of Fig.~\ref{fig:1}(a) we add another spin domain $C$ collectively coupled with domain $B$ to a second reservoir. That is, domains $A$ and $B$ are collectively coupled to the same \cjosephine{zero-temperature} reservoir and domains $B$ and $C$ are collectively coupled to a different \cjosephine{zero-temperature} reservoir. \cjosephine{This means we have the simultaneous action of two collective dissipators, $J_A^-+J_B^-$ and $J_B^-+J_C^-$.} For an initial state of 
\begin{equation}
\ket{\psi}_{AB} = \ket{\uparrow\uparrow\cdots\uparrow}_A\otimes \ket{\downarrow}_B \otimes \ket{\downarrow}_C
\end{equation}
we can expect that for $N_A$ sufficiently large, the coupling to the first reservoir will result in the spins in $A$ relaxing to the ground state and the single spin in $B$ will move to its excited state. For large $N_A$, domains $A$ and $B$ will relax at a time that scales with ${1}/{N_A}$ due to the superradiant decay \cite{dicke1954coherence,gross1982superradiance}. We can then consider the coupling to the second reservoir. In this case, the single spin in $B$ will be excited (for large $N_A$) and the single spin in $C$ has already been initialised in its ground state. With the collective coupling to the second reservoir, we can expect these two spins to end in an entangled state. In fact, for $N_A\to \infty$, we can expect the relaxation of $B$ and $C$ to proceed to the steady state described in equation \eqref{eq:0}  which is an entangled state with entanglement of formation $E_F=0.354$, that is $0.354$ ebits. 

 In this work we explore the generation of entanglement through multiple spin domains with multiple reservoir couplings. Unlike previous works investigating entangling qubits through interaction with a common reservoir \cite{hormeyll2009environment, braun2002creation,kim2002entanglement, oh2006entanglement, contreraspulido2008entanglement, maniscalco2008protecting,francica2009resonant,lin2013dissipative,shankar2013autonomously}, we will show here generation of entanglement between two spins in separate domains which have never interacted and are not even coupled to the same reservoir. In particular, we will also see the amount of entanglement (between qubits not coupled to the same reservoir) to be of similar level to that of the previously mentioned example of two qubits collectively coupled to the same zero temperature reservoir. This paper is arranged in the following way: In Sec.~\ref{sec:model} we present details of the model considered in this work. In Sec.~\ref{sec:dynamics}, we use that model to analyse the capabilities of our system to distribute entanglement. We then look at how entanglement distribution is affected by realistic conditions such as individual decay, dephasing and thermal reservoirs in Sec.~\ref{sec:dep}. Finally, we will also show how multiple system-reservoir couplings can be used to generate tripartite entangled states in Sec.~\ref{sec:trip}, before presenting our summary and conclusions in Sec.~\ref{sec:summary}. 

\section{Our model \label{sec:model}}
The system we examine in this work consists of $M$ separate spin domains, each consisting of $N_m$ identical spin-1/2 particles with frequency $\omega_0/2\pi$. Although in our system, we assume each spin has the same energy, entanglement can still be generated through a common environment when this is not the case \cite{auyuanet2010quantum}. The spins in two neighbouring spin domains are collectively coupled to a bosonic reservoir. Additionally, our system is symmetric under exchange of two spin-1/2 particles in a given spin domain. The Hamiltonian for our system of $M$ spin domains and $M-1$ reservoirs is given by:
\begin{align}
&H=\hbar \omega_0  \sum_m^M J_m^z  +\sum_m^{M-1} \sum_{k_i} \hbar \omega_{k_i} a_{k_i}^\dagger a_{k_i}  \nonumber
\\
&+\sum_m^{M-1}\sum_{k_i}  \left[ t_{k_m} \left(J_m^+ +J_{m+1}^+ \right)a_{k_m}  + t_{k_m}^* a_{k_m}^\dagger \left(  J_m^-+J_{m+1}^- \right) \right]
\label{eq:ham}
\end{align}
where the operators $a^\dagger_{k_m}$ ($a_{k_m}$) are the creation (annihilation) operators of the ${k_{m}^{\mathrm{th}}}$ bosonic mode of the reservoirs with frequency $\omega_{k_m}/2\pi$. The operators $J_i^\alpha$ are the collective spin operators for the $m^{\mathrm{th}}$ spin domain and are given by $J_m^\alpha= 1/2 \sum_{n_m=1}^{N_m} \sigma_{n_m}^\alpha $ where $\sigma^\alpha_{n_m}$ are the $n_m^{\mathrm{th}}$ Pauli spin operators with $\alpha=x,y,z$. The collective spin raising and lowering operators for the $m^{\mathrm{th}}$ spin domain are $J_m^{\pm}=J_m^x \pm i J_m^y$. In~\eqref{eq:ham}, the first term represents the Hamiltonian of the spin domains, the second term represents the Hamiltonian of the bosonic reservoirs and the third term represents the interaction between the spin domains and the reservoirs. In the third term, each term in the sum over $M-1$ reservoirs describes the coupling between the $m^{\mathrm{th}}$ and  $\left(m+1\right)^{\mathrm{th}}$ spin domain which are collectively coupled to the $m^{\mathrm{th}}$ bosonic reservoir, with $t_{k_m}$ ($t_{k_m}^*$) representing the emission (absorption) amplitudes that fix the spectral density of the reservoirs, $\Gamma_m\left( \omega \right)= 2\pi \sum_{k_m} |t_{k_m}|^2 \delta \left( \omega -\omega_{k_m} \right) $.  Within the Born-Markov-secular approximation, the master equation in the rotating frame can be written as \cite{gorini2008completely, lindblad1976generators,BreuerPetruccioneBook2002,carmichael2013statistical}:
\begin{align}
\begin{split}
\label{eq:master}
\dot{\rho}_s=  \sum\limits_i^{M-1}\frac{\gamma_{m}}{2}\mathcal{D}\left[J_m^-+J_{m+1}^-\right]\rho_s
\ ,
\end{split}
\end{align}
for bosonic reservoirs at zero temperature with $\mathcal{D} \left[O \right] \rho=2 O \rho O^\dagger -O^\dagger O \rho -\rho O ^\dagger O $. Previous work has experimentally implemented a double spin domain system coupled to a common resonator where each spin domain consisted of nitrogen-vacancy (NV) ensembles in diamonds \cite{astner2017coherent}. \cjosephine{Additionally, the work of Ref.~\cite{ fauzi2021double} implemented a double nuclear spin-domain system coupled to the same bosonic mode. The nuclear spin ensemble size is on the order of $N>10^{12}$ which means that the time scale on which collective effects like superradiant decay occur is extremely fast ($<10^{-10}\,\mathrm{s}$) while the associated individual dephasing and thermalisation times are much slower (on the order of $T_2^*\sim 1\,\mathrm{ms}$ and $T_1\sim 40\,\mathrm{s}$ respectively). }
In this work, we primarily consider the case of three spin domains coupled to two reservoirs and therefore our master equation is given specifically by:
\begin{equation}
\dot{\rho}_s = \frac{\gamma}{2} \mathcal{D} \left[ J_A^-+J_B^-\right]\rho_s+\frac{\gamma}{2} \mathcal{D} \left[ J_B^-  + J_C^- \right] \rho_s
\label{eq:masterABC}
\end{equation}
where we have used the same reservoir coupling constant $\gamma$ for all system-reservoir couplings. Further, we can introduce a scaled time $\gamma t/2$ to simplify our considerations.

\section{Entanglement dynamics \label{sec:dynamics}}
In this work, we extend the results of Ref.~\cite{hama2018negative} to more than two spin domains with more than one reservoir coupling. We will show here that entanglement can be generated not just through collective coupling to the same reservoir but also through a chain setup where entanglement will be generated between spins which are not coupled to the same reservoir. 

\begin{figure}
	\centering
	\includegraphics[width=0.83\linewidth]{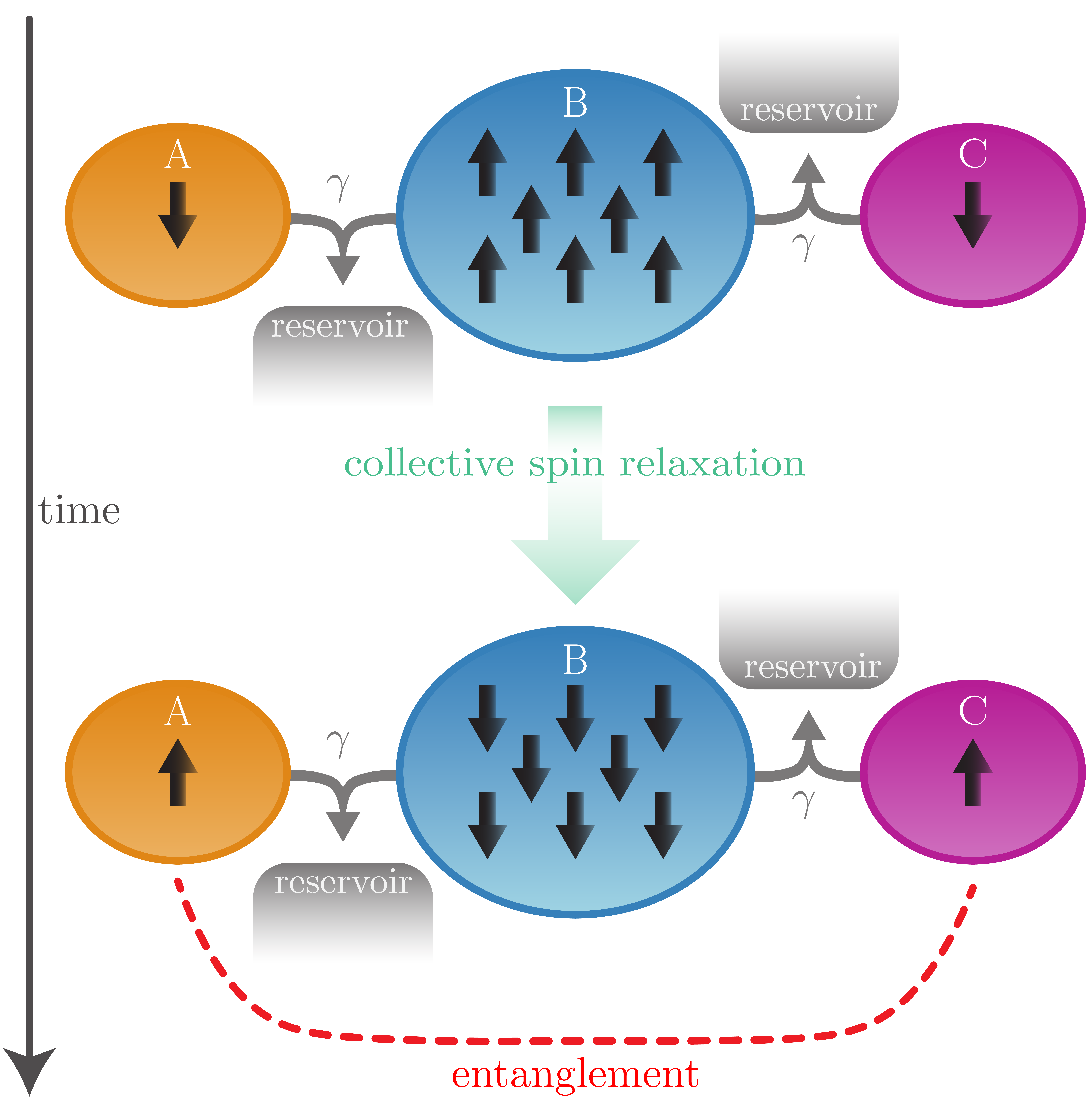}
	\caption{Three spin domains pairwise coupled to two separate reservoirs at zero temperature. Collective relaxation of the spins produces entanglement between the spins. Entanglement between spins in outer domains is maximised for a configuration with a large number of spins in the middle domain.}
	\label{fig:2}
\end{figure} 
To begin, we will conisder here the situation where domains $A$ and $C$ consist of a single spin in each domain, while the central domain $B$ has $N_B$ spins. This is depicted in Fig.~\ref{fig:2}. Our goal is to entangle the qubits (in domains $A$ and  $C$) which are not coupled to the same reservoir. After this sytem evolves in time according to equation \eqref{eq:masterABC}, through collective spin relaxation, entanglement is generated between all the spins, however, here we primarily focus on the entanglement between the two single spins in domains $A$ and $C$. In the remainder of this paper, we will examine how to maximise the entanglement between spins in $A$ and $C$ as well as how robust this entanglement is to realistic imperfections.

Our initial state configuration is:
\begin{equation}
\ket{\psi_{\mathrm{in}}} = \ket{\downarrow}_A \otimes \ket{\uparrow \cdots \uparrow}_B \otimes \ket{\downarrow}_C 
\label{eq:init}
\end{equation}
which is depicted in the top part of Fig.~\ref{fig:2}.  However, different initial state configurations can still produce entanglement (see \hyperref[sec:app]{Appendix~\ref{sec:app}}). After our initial state~\eqref{eq:init}, evolves in time we then trace out the spins in domain $B$ leaving us with the reduced density matrix $\rho_{AC}=\mathrm{Tr}_B \left[ \rho\right] $ of the state of the two spins, one in domain $A$ and one in domain $C$. To measure the amount of entanglement present in the state $\rho_{AC}$, we will use the entanglement of formation, which quantifies the minimum amount of pure state entanglement needed to prepare the state \cite{bennett1996purification}. For an arbitrary state of two qubits, this quantity is given by \cite{hill1997entanglement}:
\begin{equation}
E_F\left( {\rho} \right) =\mathcal{E} \left(C\left(\rho\right) \right)
\end{equation}
 where $\mathcal{E}$ is given by $\mathcal{E}\left(C\right) =h\left(\left[1 +\sqrt{1-C^2}\right]/2\right)$  with $h\left(x\right)=-x \log_2 x -\left(1-x\right) \log_2 \left(1-x\right)$. The concurrence $C$ is given by $C\left(\rho\right) =\max\{ 0, \lambda_1-\lambda_2-\lambda_3-\lambda_4 \}$ where $\lambda_i$ are the square roots of the eigenvalues of the matrix $\rho \tilde{\rho}$ with $\tilde{\rho}=\left(\sigma_y\otimes \sigma_y\right)\rho^* \left(\sigma_y\otimes \sigma_y\right)$ \cite{wootters1998entanglement}.

\begin{figure}
	\centering
	\includegraphics[width=0.99\linewidth]{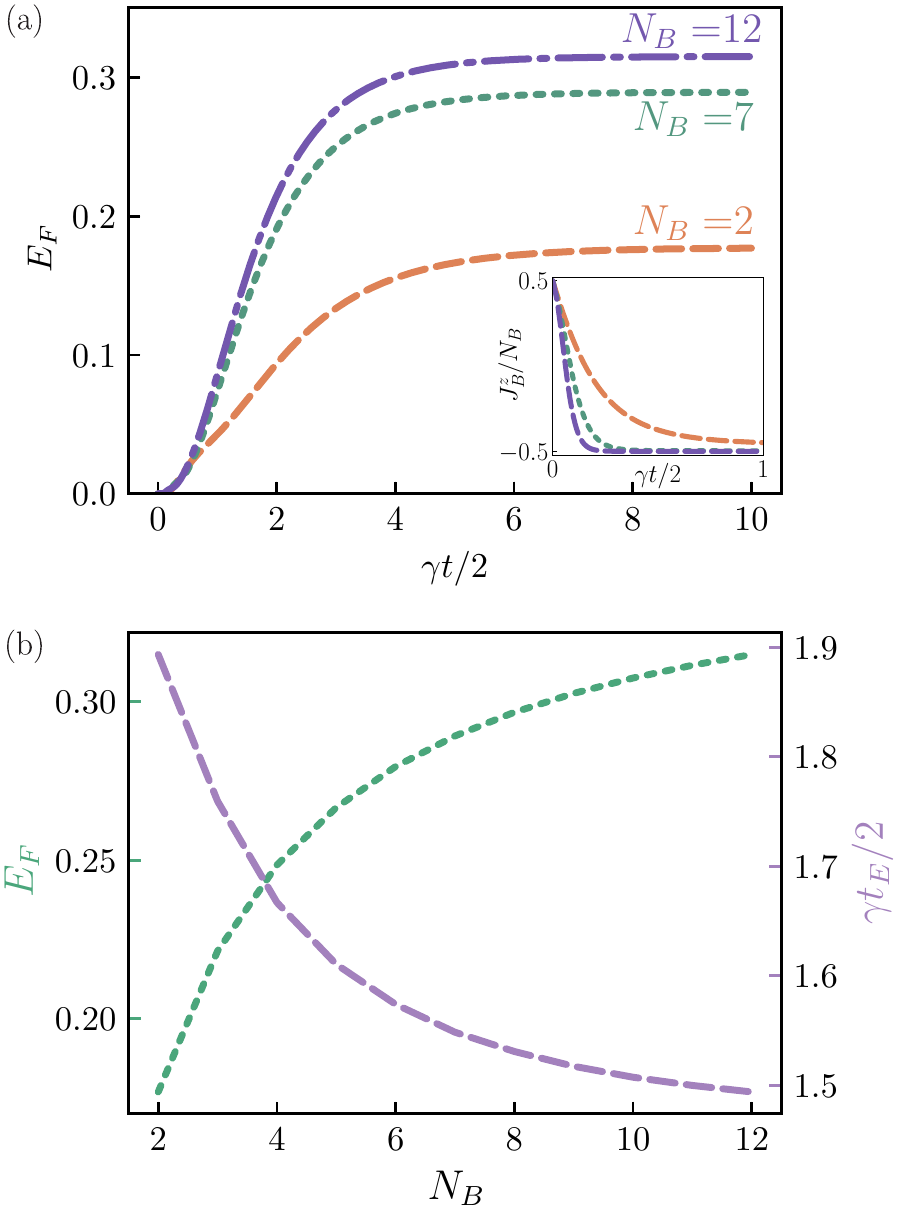}
	\caption{Entanglement generated between spins in $A$ and $C$ using the configuration shown in Fig.~\ref{fig:2} . (a) The entanglement of formation $E_F$ versus the scaled time $\gamma t/2$ for various values of $N_B$ (the spin-population in domain $B$) which clearly shows the entanglement generation is higher and faster for large $N_B$. Further, shown in the inset is the normalised collective spin relaxation $J^z_B/N_B$ illustrating the enhanced decay dynamics due to superradiance \cite{dicke1954coherence,gross1982superradiance}. (b) Steady state entanglement $E_F$ and speed of entanglement generation  $\gamma t_E/2$ as a function of the spin-population $N_B$. The left axis shows the steady state entanglement of formation $E_F$ generated between spins in $A$ and $C$ (green, dotted line). The right axis shows the value of $\gamma t_E/2$ where $t_E$ is the time it takes for entanglement between $A$ and $C$ to reach half of its maximum value (purple, dashed line).}
	\label{fig:eof-jz}
\end{figure} 
In Fig.~\ref{fig:eof-jz}(a) we plot the entanglement generation dynamics between the two spins in domains $A$ and $C$ as well as the relaxation dynamics of the initially excited domain $B$. Here we can see that as the number of spins in domain B increases, the decay time decreases. This is indicative of the collective effect of superradiance \cite{dicke1954coherence,gross1982superradiance} where the radiation is enhanced and sped up in the presence of mutliple emitters. The increase in spin domain size also contributes to enhanced entanglement dynamics - that is the speed of entanglement generation is increased as is the amount of entanglement present between qubits in $A$ and $C$. Note that only collective dissipation into a zero temperatature reservoir is govering the dynamics, and for this case, the entanglement present in the steady state remains and does not decay. We also emphasise here that entanglement is generated  between spins which are not coupled to the same reservoir. For the system size of $N_B=12$, the entanglement generated between spins in domains $A$ and $C$ is $E_F\approx 0.315$ ebits. 

Next in Fig.~\ref{fig:eof-jz}(b), we plot the maximum value of the entanglement of formation $E_F$ (reached at steady state) as well as the value of $\gamma t_E /2$ where $t_E$ is the time for entanglement between $A$ and $C$ to reach half of the maximum entanglement present during the system evolution. It can clearly be seen that the entanglement distribution is enhanced (more entanglement and faster distribution) for a large spin domain size in $B$. Results here are limited to small total system sizes ($N_A+N_B+N_C$) due to the memory required to simulate the combined system. As can be seen from Fig~\ref{fig:eof-jz}(b) as we increase $N_B$,  the increase in entanglement is rapid at first, however the growth steadily slows. Thus, for even bigger $N_B$ the payoff with increased entanglement between $A$ and $C$ will not be significant. \cjosephine{We find that for $N_B\to \infty$, the steady state of the reduced system $\rho_{AC}$ goes to 
\begin{equation}
\rho_{AC} =\frac{1}{2} \ket{\downarrow\downarrow}\bra{\downarrow\downarrow} +\frac{1}{2} \ket{\Psi^+} \bra{\Psi^+}
\end{equation}
where $\ket{\Psi^+}=1/\sqrt{2} \left(\ket{\uparrow\downarrow}+\ket{\downarrow\uparrow}\right)$, and thus the entanglement of formation of this state is the same as the first example in the introduction, that is $\mathcal{E}\approx0.354 \,\mathrm{ebits} $. In \hyperref[app:steadystate]{Appendix~\ref{app:steadystate}}, we also illustrate the steady state solution for any spin population $N_B$  with initial state $\ket{\frac{1}{2}}_A \otimes  \ket{\text{-}\frac{N_B}{2}}_B\otimes \ket{\text{-}\frac{1}{2}}_C$.
}

\begin{figure}
	\centering
	\includegraphics[width=0.9\linewidth]{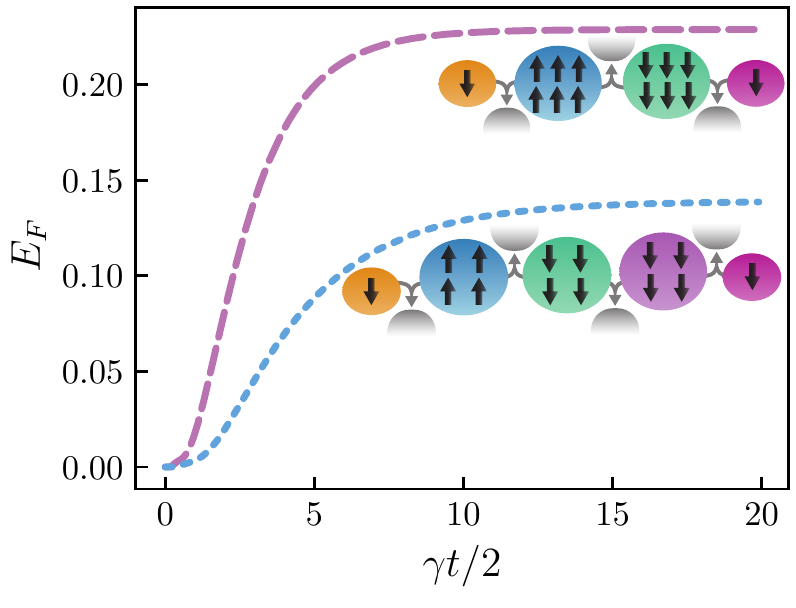}
	\caption{The entanglement of formation $E_F$ generated between spins in the outer domain of a spin domain chain versus the scaled time $\gamma t/2 $. The pink (dashed) line shows $E_F$ for spins in the outer domains of a four spin domain chain connected with three reservoirs, depicted in the inset shown in the upper right corner. The spin domain population configuration for this result is $\{ N_A, N_B, N_C, N_D \}=\{ 1,6,6,1\}$. The blue (dotted) line shows the $E_F$ for spins in the outer domains of a five spin domain chain with four reservoirs, depicted in the inset diagram below the blue (dashed) line. The spin domain population configuration for this result is $\{ N_A, N_B, N_C, N_D, N_E \}=\{ 1,4,4,4,1\}$. }
	\label{fig:eof45}
\end{figure} 
While the previous results focus on distributing entanglement between spins in the outer domains of a three spin domain chain, entanglement generation through collective spin relaxation is not limited to three domains. In Fig.~\ref{fig:eof45} we present results for the entanglement between the single spins in the outer domains of spin domain chains consisting of four (pink, dashed line) and five (blue, dotted line) spin domains. These dynamics use the initial state depicted in the inset diagrams in Fig.~\ref{fig:eof45} respectively, that is all spins initialised in the ground state with the exception of domain $B$ which has all spins initialised in the excited state. While we only show entanglement for a single spin domain population configuration for four and five domains, as before, entanglement between outer qubits is maximised for larger spin populations of the middle domains. Here, we have constrained the total number of spins in all middle domains to be the same (12 spins), however the top line uses 12 spins split between two spin-domains and the bottom line uses 12 spins split between three spin-domains. These results are limited due to the total size of the system however, they show that entanglement can be distributed between qubits separated by two or three spin domains which are in turn connected by collective coupling to three or four reservoirs.  For this result, as is the case with the previous results, the only effect taken into account in this model is collective decay into zero-temperature reservoirs. In this case, we observe constant entanglement which does not decay in the steady state.

\section{Individual decay and dephasing \label{sec:dep}}
\begin{figure*}
	\centering
	\includegraphics[width=0.99\linewidth]{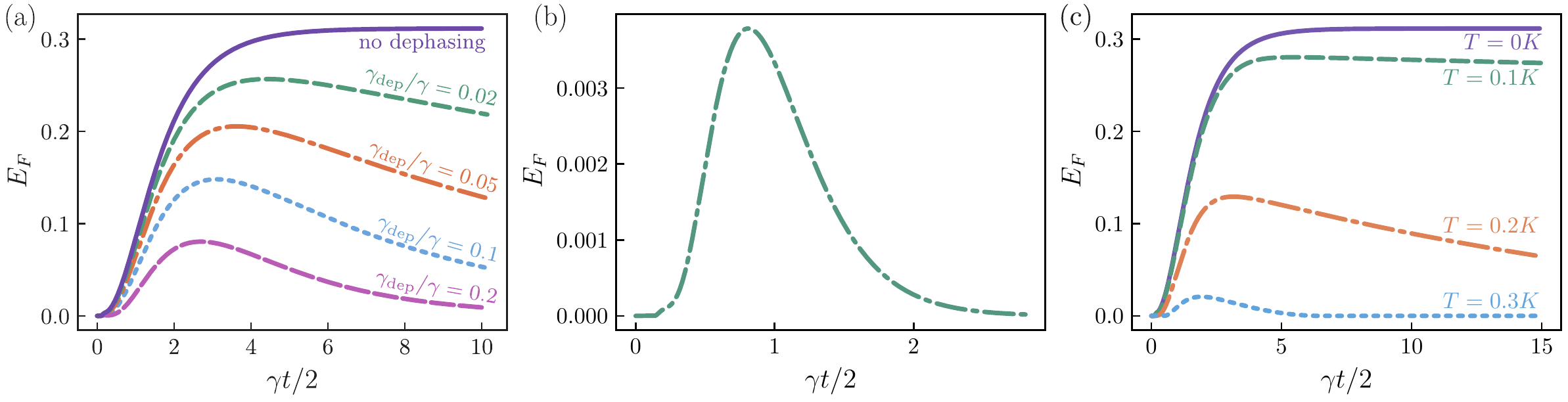}
	\caption{The entanglement of formation $E_F$  generated between spins in domains $A$ and $C$  versus the scaled time $\gamma t/2 $  with a spin population of $N_B=11$ in domain $B$. In (a), we show various dephasing rates $\gamma_{\mathrm{dep}}$.  The pink (long dashed) line shows the effect of dephasing with $\gamma_{\mathrm{dep}}/\gamma=0.2$, the blue (dotted) line uses $\gamma_{\mathrm{dep}}/\gamma=0.1$, the orange (dash-dot) line uses $\gamma_{\mathrm{dep}}/\gamma=0.05$, the green (dashed) line uses $\gamma_{\mathrm{dep}}/\gamma=0.02$, and the purple (solid) line shows the ideal case of $\gamma_{\mathrm{dep}}/\gamma=0$. In (b), we show the entanglement dynamics with both collective and individual decay. In (c), we use a thermal reservoir of temperature $T$ with the following parameters chosen: $\omega_0/2\pi=10\, \mathrm{G Hz}$ and $\gamma =0.01\, \mathrm{Hz}$.}
	\label{fig:dep}
\end{figure*} 
The results contained in previous section are interesting but represent the highly specific and idealized system of collective relaxation with a zero-temperature bath and with no other individual effects present. It is important to consider more realistic models, which includes effects such as dephasing, individiual spin relaxation and non-zero temperature reservoirs, and examine how these aspects may alter the previously observed entanglement generation.

For the three domain system depicted in Fig.~\ref{fig:2}, we now incorporate finite temperature reservoirs as well as dephasing and individual relaxation. This yields the following master equation, again in the rotating frame of the system \cite{fauzi2021double}:
\begin{widetext}
\begin{align}
\begin{split}
\label{eq:masterdep}
\dot{\rho}_s =&  \frac{\gamma}{2} \left( \bar{n}+1\right) \left(   \mathcal{D} \left[J_A^-+J_B^-\right] \rho_s + \mathcal{D} \left[J_B^-+J_C^-\right] \rho_s  \right)
 + \frac{\gamma}{2} \bar{n} \left( \mathcal{D}  \left[ J_A^++J_B^+\right] \rho_s + \mathcal{D} \left[ J_B^+ +J_C^+\right] \rho_s \right) 
\\
&+ \sum_{i_{\mathrm{A,B,C}}=1}^{N_{\mathrm{A,B,C}}}   \frac{\gamma}{2} \left( \bar{n}+1\right)  \mathcal{D} \left[\sigma_{i_{A,B,C}}^- \right] \rho_s 
+ \frac{\gamma}{2}  \bar{n}  \mathcal{D} \left[\sigma_{i_{A,B,C}}^+ \right] \rho_s 
+ \frac{\gamma_{\mathrm{dep}}}{2}    \mathcal{D} \left[ \sigma_{i_{A,B,C}}^z\right] \rho_s
\end{split}
\end{align}
\end{widetext}
In \eqref{eq:masterdep}, the first and second terms correspond to the collective thermalisation between the pairs of spin domains ($A$ and $B$, $B$ and $C$ respectively) and the thermal reservoir. Associated with that reservoir is the Bose-Einstein distribution $\bar{n}=1/\left(e^{\hbar \omega_0/k_B T}-1\right)$ for a given temperature $T$ and with Boltzman constant $k_B$. Next, the third and fourth terms in~\eqref{eq:masterdep} correspond to individual thermalisation of the spins. Finally, the last term corresponds to dephasing. To summarise, the first line in \eqref{eq:masterdep} corresponds to collective effects while the second line corresponds to individual effects associated with thermalization and dephasing. Note that the individual and collective thermalisation occurs with the same system-reservoir coupling constant $\gamma$.

In Fig.~\ref{fig:dep}(a), we plot the entanglement dynamics of the original system with no dephasing (pink, dashed line) as well as the entanglement between spins in domains $A$ and $C$ with various dephasing rates of $\gamma_\mathrm{dep}/\gamma$ (broken lines). We observe a reduction in peak entanglement generated through the system-bath evolution in time but also a gradual reduction in the entanglement present in the steady state. For the results in Fig.~\ref{fig:dep}(a), the dephasing is the only additional effect present, the reservoirs remain at zero temeperature ($ \bar{n}=0$) and there is no individual decay. 

We then examine the entanglement generation capabilities of our system under the effects of individual spin coupling to the reservoir. In this case, the spins all have individual as well as the collective dissipation. Again, we shall keep the bosonic reservoir to zero temperature to isolate the effect of individual decay on entanglement dynamics. We find that individual coupling has a signficant detrimental effect on the entanglement present. Shown in Fig.~\ref{fig:dep}(b) is the entanglement dynamics between single spins in domains $A$ and $C$. Note the significantly smaller scale of the y-axis in Fig.~\ref{fig:dep}(b) relative to the other results in this work. The maximum entanglement of formation produced in this case is $E_F=0.0039$. Note that as the system size increases (number of spins), the superradiant collective effects occur on a much faster time scale than the individual one and most likely dephasing. For example in NV centers in diamond, the  collective relaxation can be of order of microseconds, while the individual one is 10 orders of magnitude larger \cite{angerer2018superradiant}. 

Lastly, the temperature of the thermal reservoir also substantially affects the entanglement generated through this system. Shown in Fig.~\ref{fig:dep}(c) is the entanglement dynamics between qubits in $A$ and $C$ for thermal reservoirs of various temperatures for spin domain population of $N_B=11$. For this result, we keep only collective spin-reservoir coupling to the bath. As can be seen in Fig.~\ref{fig:dep}(c), the entanglement generated through our system is highly sensitive to the temperature of the thermal reservoirs. As the temperature increases, less entanglement is generated and in the steady state entanglement is also reduced. 
 
We have seen in this section the impact of realistic effects on entanglement distribution via collective dissipation with multiple spin domains and reservoirs. While it is clear that entanglement can still be generated between spins in domains $A$ and $C$, it significantly decreases if individual decay or finite temperatures are taken into account. Further, the constant, non-decaying entanglement in the steady state observed in Sec.~\ref{sec:dynamics} is no longer present once any of these  effects are taken into account. In the results of Figs.~\ref{fig:dep}(a)-(c) we observe a decline in the entanglement of formation over time. However, individual process are always detrimental to entanglement between spins, and therefore the decay of the entanglement in the results of Fig.~\ref{fig:dep} is to be expected.

\section{Tripartite negativity \label{sec:trip}}
\begin{figure}
	\centering
	\includegraphics[width=0.80\linewidth]{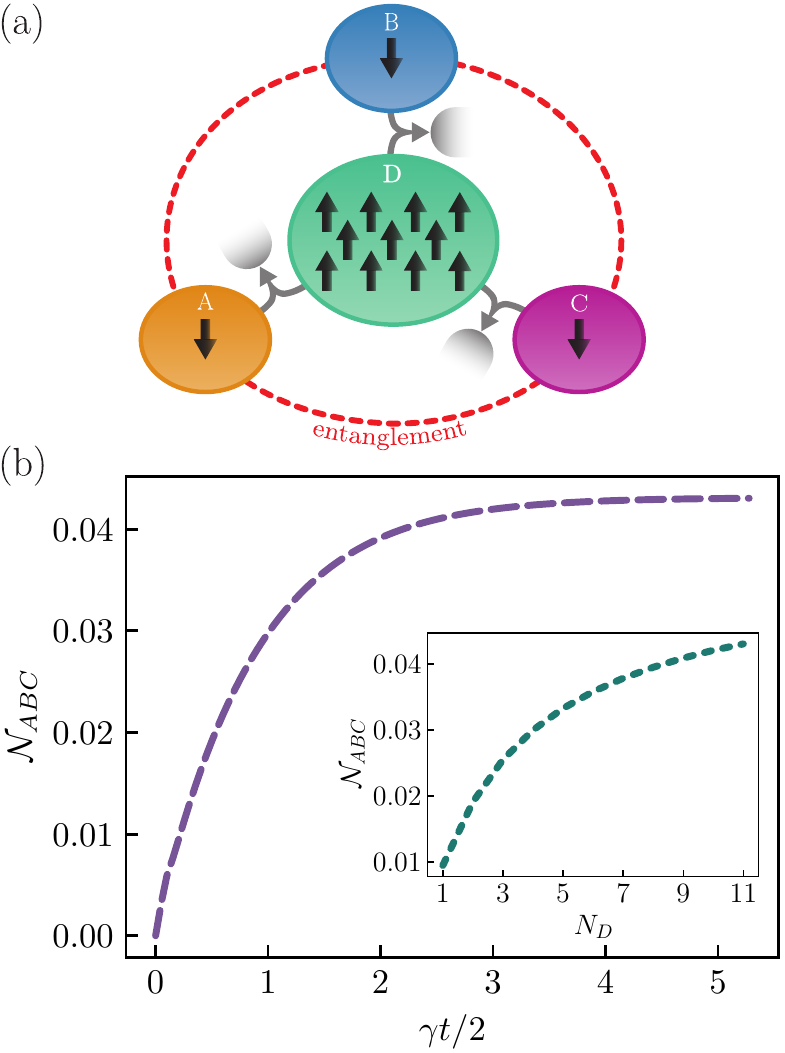}
	\caption{Tripartite quantum correlations generated from collective decay with multiple reservoirs. (a) Initially, all spins are in a separable state \eqref{eq:tri-in}  with their configuration depicted here. After collective spin relaxation, the reduced state of the three single qubits in spin-domains A, B and C are entangled in the steady state.  (b) Tripartite negativity generated between spins $A$, $B$ and $C$ during spin relaxation with the configuration pictured in (a) for a central domain populated with $N_D=11$ spins. Shown in the inset is the steady state tripartite negativity for spin population $N_D$ ranging from 1 to 11.}
	\label{fig:trip}
\end{figure} 

The previous results in this work focused solely on entangling two qubits through collective spin relaxation with multiple collective spin-reservoir couplings. In this section, we will show how this same process can be utilised to generate tripartite entanglement between three spins. Consider the setup shown in Fig.~\ref{fig:trip}(a). This set up is reminiscent of a spin-star configuration \cite{anza2010tripartite,hutton2004mediated} but we emphasise again the spins are not directly coupled, merely collectively coupled to the same reservoir as other spins. Here, a central spin-domain (labelled D) contains many spins and is collectively coupled to three bosonic reservoirs with three other spin-domains each containing a single spin only (labelled A, B and C). The system is initialised in the separable state:
\begin{equation}
\ket{\Psi}=\ket{\downarrow}_A \otimes \ket{\downarrow}_B  \otimes \ket{\downarrow}_C \otimes \ket{\uparrow \uparrow \cdots \uparrow}_D
\label{eq:tri-in}
\end{equation}
Under the process of collective dissipation with the reservoir setup illustrated by the diagram in Fig.~\ref{fig:trip}(a), the steady state of A, B and C (with domain $D$ traced out) is given by:
\begin{equation}
\rho_{ABC}  =\mathrm{Tr}_D \left[ \rho_{ABCD}\right] 
= c_G \ket{\downarrow\downarrow \downarrow} \bra{\downarrow \downarrow\downarrow} + c_W   \ket{W} 
\bra{W}
\label{eq:trss}
\end{equation}
that is, the steady state of the reduced system of three qubits A, B and C is a mixture of the ground state and the W-state which is $\ket{W}=1/\sqrt{3}\left(\ket{\uparrow \downarrow\downarrow} + \ket{\downarrow \uparrow \downarrow} + \ket{\downarrow \downarrow \uparrow} \right)$. Therefore, the steady state of this reduced system has tripartite entanglement between the three remaining qubits. In order to demonstrate the presence of tripartite entanglement in the steady state of this system, we calculate the tripartite negativitity, defined as the geometric mean of the negativities of all bipartitions  \cite{sabin2008classificationa}:
\begin{equation}
\mathcal{N}_{ABC} \left( \rho \right) = \left( \mathcal{N}_{A|BC} \mathcal{N}_{B|AC} \mathcal{N}_{C|AB} \right)^{1/3}
\label{eq:tr}
\end{equation}
where each of the terms $\mathcal{N}_{i|jk}$ are the bipartite negativities between subsystems $i$ and $jk$ given by $\mathcal{N} \left(\rho\right) =\left(||\rho^{\Gamma_i}||-1\right)/2$  with $\rho^{\Gamma_i}$ corresponding to the partial transpose of the density matrix with respect to subsystem $i$ and $||X||$ is the trace norm of the operator $X$ given by $||X||=\mathrm{Tr}\sqrt{X X^\dagger}$ \cite{zyczkowski1998volume, vidal2002computable}.  Note that with this definition of the bipartite negativity, maximally entangled Bell states have the negativity of 0.5. Further, for three qubit entangled states, GHZ states yield a tripartite negativity of 0.5 but W states have a tripartite negativity of 0.47. In Fig.~\ref{fig:trip}(b), we give the tripartite negativity generated via the configuration shown in Fig.~\ref{fig:trip}(a) with reservoirs at zero-temperature and no individual effects. The dynamics show clear presence of tripartite negativity in the steady state. For the given spin-domain population of $N_D=11$, the steady state tripartite negativity is $\mathcal{N}_{ABC}\approx0.043$.  Additionally, the inset shows the steady state tripartite entanglement for various values of $N_D$, the population of the central domain. As the spin-domain population of $D$ grows, the steady state contribution of the W-state term [$c_W$ in \eqref{eq:trss}] grows larger, and thus the tripartite negativity is maximised for large $N_D$.  However, while this increase is rapid at first, it slows as $N_D$ continues to increase indicating less payoff in tripartite correlations for further increasing the size of the central domain $D$. Nevertheless, the results here show the generation of tripartite correlations between three qubits not coupled to each other or even the same reservoir.
\section{Summary and conclusions \label{sec:summary}}
We have shown here how, by utilising reservoir engineering, and with a completely separable initial state,  entanglement can be generated between two spins which are not coupled to each other or even coupled to the same reservoir. We have focused mostly here on the case where the two spins collectively decay into different reservoirs with a single common spin domain collectively decaying into both reservoirs. For this case, the entanglement is maximised for a large number of spins in the central domain. The speed of entanglement generation is also maximised for a larger spin population in the central domain. Additionally, the entanglement betweeen the qubits not coupled to the same reservoir, can approach the level of entanglement generated when two qubits are collectively coupled to the same zero-temperature reservoir. Entanglement generation via this mechanism is not limited to the three spin-domain case, but can also occur between qubits in the outer domains of four and five domain systems as well as entangling three qubits using a large central domain of many spins.

The results in Sec.~\ref{sec:dep} show that, as expected, when realistic effects are taken into account the entanglement generation between spins suffers significantly. With dephasing or individual spin-reservoir coupling, not only is the maximum entanglement generated reduced, the entanglement in the steady state also decays over time. However, we emphasise that for large enough ensemble size, the collective effects will occur on a timescale that is much faster than that of the individual effects. Additionally, finite temperature reservoirs also reduce the maximum entanglement and cause the steady state entanglement to decay. While the results here predict rather small entanglement generation, further optimisation may significantly increase the maxium entanglement of qubits and make it more robust against imperfections.

\section{Acknowledgements}
C. W. W. acknowledges financial support by the Deutsche Forschungsgemeinschraft (DFG, German Research Foundation) – Project No. 496502542 (WA 5170/1-1). This work was supported by the JSPS KAKENHI Grants 19H00662 \& 21H04880, the MEXT Quantum Leap Flagship Program (MEXT QLEAP) Grant JPMXS0118069605 and the Moonshot R\&D Program Grant JPMJMS226C. \cjosephine{We are grateful for the help and support provided by the Scientific Computing and Data Analysis section of the Research Support Division at OIST.}

\appendix
\setcounter{equation}{0}
\section{Alternate initial states \label{sec:app}}
\begin{figure}\captionsetup{position=top}
\centering
{\includegraphics[width=0.8\linewidth]{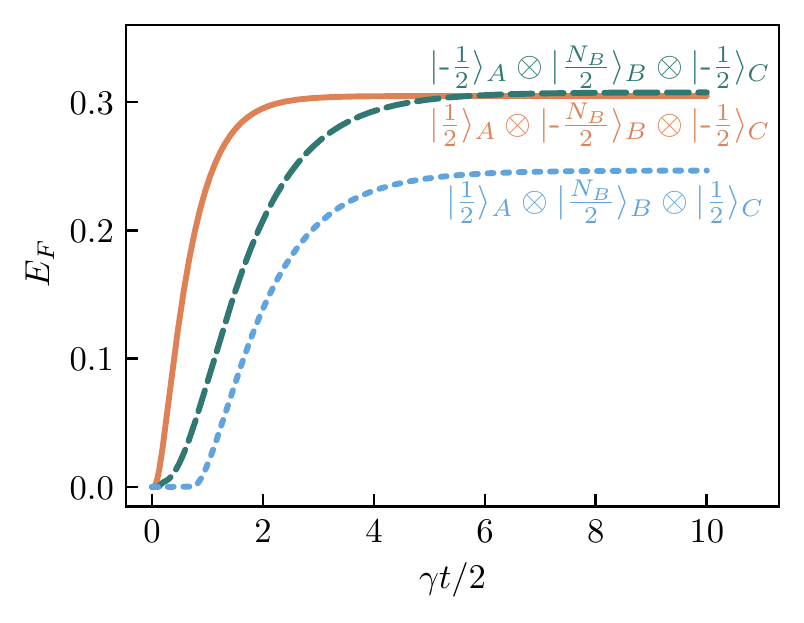}} 
\caption{\cjosephine{Entanglement of formation between single spins in $A$ and $C$ generated via collective spin relaxation \eqref{eq:masterABC} for different initial states. The various initial states are labelled in the figure. Here we have used $N_B=10$.   }}
\label{fig:init}
\end{figure}
\cjosephine{
In Sec.~\ref{sec:dynamics}, we showed the dynamics of entanglement generation in our three spin-domain system with the initial state $\ket{\psi_{\mathrm{in}}} = \ket{\downarrow}_A \otimes \ket{\uparrow \cdots \uparrow}_B \otimes \ket{\downarrow}_C $ where the middle and most populated spin domain $B$ is initialised with all spins in the excited state.  This state can also be written in the notation $\ket{\text{-}\frac{1}{2}}_A \otimes \ket{\frac{N_B}{2}}_B \otimes \ket{\text{-}\frac{1}{2}}_C$. However, this is not the only configuration that can produce entanglement through collective spin relaxation. In the following, we will explore other initial states and their effect on the steady state entanglement produced via this mechanism. }

\cjosephine{As we consider the collective coupling of many spins to the same reservoir, a necessary condition of the collective coupling is that the spins are indistinguishable from the perspective of the reservoir. Thus we limit ourselves to initial conditions that are permutationally symmetric within each spin ensemble. These states of permutational symmetry are the so-called Dicke states, and apart from the lowest and highest energy Dicke states (ground or fully excited states respectively), the Dicke states have significant entanglement \cite{garraway2011dicke,agarwal2013quantum}. In this work, our goal is to start with an initially separable state, and via reservoir engineering, result in the system in a highly entangled steady state. Thus, in exploring alternate initial states, we will limit ourselves to only the separable permutationally symmetric states, which are the fully excited state or ground state of each spin ensemble. 

In considering a three spin-ensemble system, this leaves us with two possible initial states for each of the three spin-ensembles, thus leaving eight possible configurations. We will briefly go over the outcomes for each of the eight configurations here. Firstly, with the initial state as the ground state of the entire system, $\ket{\text{-}\frac{1}{2}}_A\otimes \ket{\text{-}\frac{N_B}{2}}_B \otimes \ket{\text{-}\frac{1}{2}}_C$, this trivially produces no dynamics as it does not decay and thus remains in the fully separable ground state. In Fig.~\ref{fig:init}, we have plotted the entanglement dynamics for three different intitial states. Again, this is the entanglement of formation between the single spins in domains $A$ and $C$ after the spins in $B$ have been traced out. The dark green, dashed line shows the entanglement for the case studied in the main text of this paper, where domains $A$ and $C$, of spin populations $N_A=N_C=1$, are initialised in the ground state and the central domain, of spin population $N_B$, is initialised in the excited state.  This is indicated by the line labelled $\ket{\text{-}\frac{1}{2}}_A\otimes \ket{\frac{N_B}{2}}_B \otimes \ket{\text{-}\frac{1}{2}}_C$ on Fig.~\ref{fig:init}. On this plot, we also show the entanglement dynamics for the inital condition $\ket{\frac{1}{2}}_A \otimes  \ket{\text{-}\frac{N_B}{2}}_B\otimes \ket{\text{-}\frac{1}{2}}_C$ which is given by the solid orange line. This solid orange line also represents the entanglement dynamics of the equivalent initial condition of $\ket{\text{-}\frac{1}{2}}_A \otimes  \ket{\text{-}\frac{N_B}{2}}_B\otimes \ket{\frac{1}{2}}_C$. Thus, three different initial conditions all produce strikingly similar entanglement levels in the steady state.  }  

\cjosephine{The strong entanglement in the steady state reached from the initial condition $\ket{\frac{1}{2}}_A\otimes \ket{\frac{N_B}{2}}_B \otimes \ket{\frac{1}{2}}_C$ (blue, dotted line in Fig.~\ref{fig:init}) highlights the novelty of the system considered in this work. To explain this, consider the following: If the three spin ensembles $A$, $B$ and $C$ were all coupled to a single reservoir, under the collective dissipator $J_A^-+J_B^-+J_C^-$, the system would relax to the fully separable ground state of the entire system. As can be seen on Fig.~\ref{fig:init}, the steady state from this initial condition is highly entangled. This means that, as the system relaxes, a significant proportion of the state remains in an entangled dark state and this is purely due to the collective coupling to \emph{two} zero-temperature reservoirs, that is the simultaneous action of the dissipators $J_A^-+J_B^-$ and $J_B^-+J_C^-$. 
}

\cjosephine{The remaining initial conditions of $\ket{\frac{1}{2}}_A \otimes  \ket{\text{-}\frac{N_B}{2}}_B\otimes \ket{\frac{1}{2}}_C$ and $\ket{\frac{1}{2}}_A \otimes  \ket{\frac{N_B}{2}}_B\otimes \ket{\text{-} \frac{1}{2}}_C$ with its equivalent $\ket{\text{-}\frac{1}{2}}_A \otimes  \ket{\frac{N_B}{2}}_B\otimes \ket{ \frac{1}{2}}_C$ produce very little entanglement in the steady state. All three initial states have less than $0.006$ ebits of entanglement of formation.  Thus, we can say that, for three spin ensembles, of the eight possible initial state configurations, four of these configurations produce significant steady state entaglement. 
}
\begin{figure}\captionsetup{position=top}
\centering
{\includegraphics[width=0.8\linewidth]{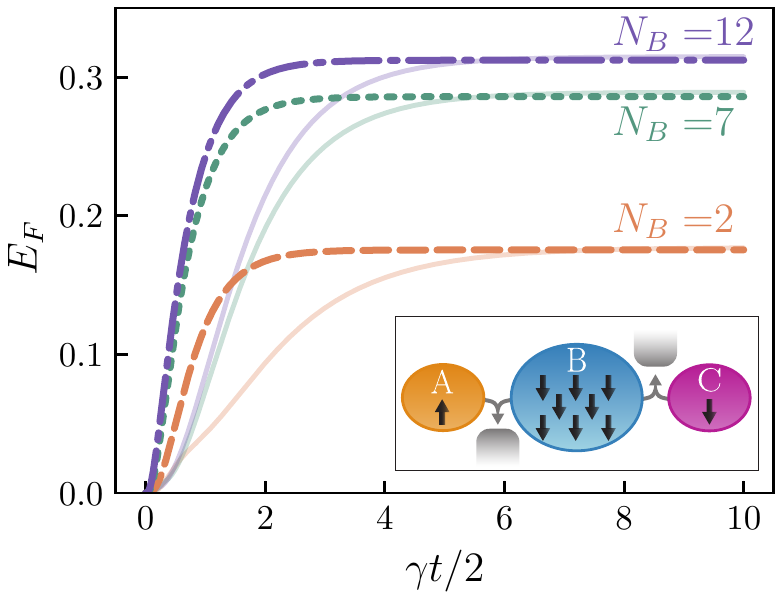}}
\caption{\cjosephine{Entanglement of formation generated via collective spin relaxation using the initial state where the single spin in domain $A$ is in its excited state while the spins in domains $B$ and $C$ are in their ground state. This state given in \eqref{eq:alt-is} is depicted in the inset. The generated entanglement of formation $E_F$ versus scaled time $\gamma t/2$ is given for $N_B=2,7,12$. We clearly observe an increase in $E_F$ increases as $N_B$ increases. The transparent, solid lines show the entanglement dynamics of the initial state \eqref{eq:init} (that is the result shown in Fig.~\ref{fig:eof-jz})  and these results are given for the same $N_B$ which is via the corresponding color.  }}
\label{fig:alt-is}
\end{figure}

\cjosephine{In Fig.~\ref{fig:alt-is}, we  take a closer look at the entanglement dynamics for the initial state
\begin{equation}
\ket{\psi_{\mathrm{in}}} =\left| \frac{1}{2} \right\rangle_A \otimes  \left| \text{-}\frac{N_B}{2} \right\rangle _B\otimes \left| \text{-}\frac{1}{2} \right\rangle_C ,
\label{eq:alt-is}
\end{equation}
and use this initial state to reproduce the result of Fig.~\ref{fig:eof-jz}(a).  As was already observed in Fig.~\ref{fig:init}, the maximum entanglement generated from intial state \eqref{eq:alt-is} is remarkably similar albeit slightly lower than that generated from initial state \eqref{eq:init}. This can be seen in Fig.~\ref{fig:alt-is} by comparing the dotted, dashed and dash-dotted lines with their corresponding solid, transparent line of the same color. The system reaches the maximum entanglement slightly faster for initial condition \eqref{eq:alt-is} than  \eqref{eq:init}.  Additionally, we find again here, an enhancement in entanglement dynamics, that is faster and more entanglement in the steady state for a larger number of spins $N_B$. However, it is worth emphasising that the number of excitations in the initial state for all lines in Fig.~\ref{fig:alt-is} is fixed at 1. This indicates that it is not the collective, superradiant decay of many spins that enhances the entanglement dynamics, but simply the presence of many spins interacting collectively with the same reservoirs.
}

\begin{figure}\captionsetup{position=top}
\centering
{\includegraphics[width=0.9\linewidth]{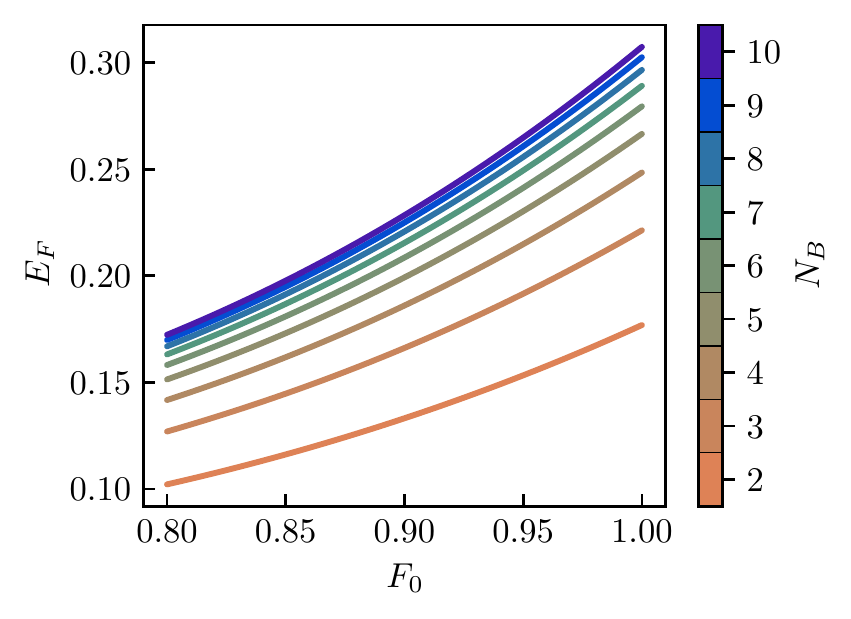}} 
\caption{\cjosephine{Steady state entanglement of formation between single spins in $A$ and $C$ for a mixed initial state. The initial state used here is given in \eqref{eq:mix}. Shown here is the steady state entanglement for a range of spin-ensemble $B$ population $2\leq N_B\leq10$ (indicated via color) and initial state fidelities $F_0$.}}
\label{fig:mix}
\end{figure}
\cjosephine{Finally, we ask what is the effect of mixed initial state preparation on the steady state entanglement produced in this scheme. As all other results in this work use pure initial states, it is wothwhile to examine whether imperfect state preparation can be detrimental to the steady state entanglement. To answer this question, we return to the main initial condition examined in this work $\ket{\text{-}\frac{1}{2}}_A\otimes \ket{\frac{N_B}{2}}_B \otimes \ket{\text{-}\frac{1}{2}}_C$ and consider the case where the preparation of the central domain $B$ is imperfect and a mixed state is prepared instead. We can model this by the following state
\begin{equation}
\rho_B = a \ket{\uparrow \uparrow \cdots \uparrow}\bra{\uparrow\uparrow\cdots \uparrow}+b \, I,
\label{eq:rhob}
\end{equation}
that is, a mixture of the target state $\ket{\uparrow \uparrow \cdots \uparrow}$ and the maximally mixed state, given by the identity matrix $I$. We assume that  preparation of the ground state can be done with high fidelity, thus the intial state of the total system is given by:
\begin{equation}
\rho_{\mathrm{in}}= \ket{\downarrow}_A\bra{\downarrow}_A \otimes \rho_B \otimes \ket{\downarrow}_C\bra{\downarrow}_C,
\label{eq:mix}
\end{equation} 
with $\rho_B$ defined as in \eqref{eq:rhob}. In Fig.~\ref{fig:mix}, we give the steady state entanglement of formation between the single spins in $A$ and $C$ for a given initial state fidelity $F_0$. To be precise, $F_0$ is the fidelity of the initial state $\rho_B$ with the target state $\ket{\uparrow\uparrow\cdots\uparrow}_B$, that is $F_0=\braket{\uparrow\uparrow\cdots \uparrow |\rho_B|\uparrow\uparrow\cdots \uparrow}$. On Fig.~\ref{fig:mix}, we can see that while a decrease in initial state fidelity does have a detrimental impact on the steady state entanglement, the scheme still produces significant entanglement even when the initial state is not completely pure. 
}

\section{Steady state and entanglement behavoir for $N_B \to \infty$ \label{app:steadystate}}
\cjosephine{
In this section, we illustrate how the steady state of our system for any $N_B$ (population of the central spin domain) can be found for the specific given initial condition \eqref{eq:alt-is}. 
To begin, we acknowledge that through numerical simulation we have found that for a system initialised in the state:
\begin{equation}
\ket{\psi_{\mathrm{in}}} =\ket{\uparrow}_A \otimes \ket{\downarrow\downarrow\cdots \downarrow}_B \otimes \ket{\downarrow}_C
\end{equation}
the steady state reached under collective spin relaxation according to master equation \eqref{eq:masterABC} is given by:
\begin{equation}
\begin{split}
\rho_{\mathrm{out}} = \left(1-x_{\mathrm{d}}\right) \ket{\downarrow\downarrow\cdots\downarrow \downarrow}\bra{\downarrow\downarrow\cdots\downarrow\downarrow}
+ x_{\mathrm{d}} \ket{\psi_{\mathrm{d}}} \bra{\psi_{\mathrm{d}}}
\end{split}
\label{eq:steadyNb}
\end{equation}
the first term is the ground state of the total system and the second term $\ket{\psi_\mathrm{d}}$ is a dark state that does not decay under either dissipator $J_A^-+J_B^-$ or $J_B^-+J_C^-$. While such a dark state is not unique, for our initial state, the form of this particular dark state is always:
\begin{equation}
\begin{split}
&\ket{\psi_{\mathrm{d}}}=  \mathcal{N} \Bigg\{
\ket{\uparrow}_A  \ket{\downarrow\cdots\downarrow}_B  \ket{\downarrow}_C
+
\ket{\downarrow}_A  \ket{\downarrow\cdots\downarrow}_B  \ket{\uparrow}_C
\\
&-\frac{1}{N_B}  \bigg[ \ket{\downarrow}_A 
 \underbrace{  \left( \ket{\uparrow \downarrow\cdots \downarrow}_B + \cdots +\ket{\downarrow\cdots \downarrow\uparrow}_B \right) }_{\text{all permutations of a single spin up in }B }  \ket{\downarrow}_C
\bigg]
\Bigg\}
\end{split}
\label{eq:darkNb}
\end{equation}
where $\mathcal{N}$ is needed for normalisation and is given by 
\begin{equation}
\mathcal{N}=\sqrt{\frac{N_B}{2 N_B+1}} \, .
\end{equation} We stress that all numerical simulations (which have been performed for $1\geq N_B\geq 12$) have yielded the steady state as \eqref{eq:steadyNb} with the given dark state \eqref{eq:darkNb}. Thus, while we know the dark state for any $N_B$, to fully characterise the steady state we still need to find the dependence of $x_{\mathrm{d}}$ on $N_B$, that is the proportion of the state that remains entangled and in the dark state as the system relaxes to its steady state. 

To find this dependence of $x_{\mathrm{d}}$ on $N_B$ we start by examining a simple case of $N_A=1$, $N_B=2$ and $N_C=1$. This system is initialised in the state
\begin{equation}
\ket{\psi_{\mathrm{in}}} =\ket{\uparrow}_A  \ket{\downarrow\downarrow}_B \ket{\downarrow}_C
\label{eq:alt2}
\end{equation}
and will relax to a steady state of the form
\begin{equation}
\rho_{\mathrm{out}}= \left(1-x_{\mathrm{d}}\right) \ket{\downarrow\downarrow\downarrow\downarrow}\bra{\downarrow\downarrow\downarrow\downarrow} + x_{\mathrm{d}} \ket{ \psi_{\mathrm{d}}}\bra{\psi_{\mathrm{d}}}, 
\label{eq:rhoxd}
\end{equation}
where for $N_B=2$, this dark state is:
\begin{equation}
\ket{\psi_{\mathrm{d}}} = \sqrt{\frac{2}{5}} \left( \ket{\uparrow\downarrow\downarrow\downarrow} -\frac{1}{2} \ket{\downarrow\uparrow\downarrow\downarrow}-\frac{1}{2}\ket{\downarrow\downarrow\uparrow\downarrow}+\ket{\downarrow\downarrow\downarrow\uparrow}\right)
\label{eq:dark2}
\end{equation}
We now define two new states, and, as the initial state \eqref{eq:alt-is} only contains a single excitation, we only need to be concerned with the single excitation subspace of the total system. For convenience, we choose the two states to be the following:
\begin{equation}
\begin{split}
\ket{\psi_1} &=  \frac{1}{\sqrt{10} }\left(  \ket{\uparrow\downarrow\downarrow\downarrow} +2  \ket{\downarrow\uparrow\downarrow\downarrow}+ 2\ket{\downarrow\downarrow\uparrow\downarrow}+\ket{\downarrow\downarrow\downarrow\uparrow} \right)
\\
\ket{\psi_2 } &= \frac{1}{\sqrt{2} } \left(  \ket{\uparrow\downarrow\downarrow\downarrow}  - \ket{\downarrow\downarrow\downarrow\uparrow}  \right)
\end{split}
\end{equation}
These states were chosen as it can be easily verified that a system starting in initial state $\ket{\psi_1}$ or $\ket{\psi_2}$ and evolving under the master equation \eqref{eq:masterABC},  will relax to the ground state of the total system $\ket{\downarrow\downarrow\downarrow\downarrow}$.  Note that the states $\ket{\psi_1}$, $\ket{\psi_2}$ and $\ket{\psi_{\mathrm{d}}}$ are all mutually orthogonal and if we wanted to define a basis to span the entire single excitation subspace, the remaining basis state would be $\ket{\psi_{\mathrm{d}_2}}= {1}/\sqrt{2}\left( \ket{\downarrow\uparrow\downarrow\downarrow}-\ket{\downarrow\downarrow\uparrow\downarrow}\right)$ which is also a dark state under both dissipators $J_A^-+J_B^-$ or $J_B^-+J_C^-$. 
However, we only need $\ket{\psi_1}$, $\ket{\psi_2}$ and $\ket{\psi_{\mathrm{d}}}$ as the initial state \eqref{eq:alt2} can be written in terms of these states:
\begin{equation}
\begin{split}
\ket{\psi_{\mathrm{in}}} &=\ket{\uparrow}_A  \ket{\downarrow\downarrow}_B \ket{\downarrow}_C \\
&= \frac{1}{\sqrt{10}} \left( \ket{\psi_1} + \sqrt{5}\ket{\psi_2}+\sqrt{4} \ket{\psi_{\mathrm{d}}}\right)
\label{eq:in-altbasis}
\end{split}
\end{equation}
In a similar fasion as the example in the introduction, we can now write down the form of the steady state because we know that the first two terms in \eqref{eq:in-altbasis} will relax to the ground state while the last term is a dark state and does not decay. Thus, the steady state of the total system is:
\begin{equation}
\rho_{\mathrm{out}} =  \frac{3}{5} \ket{\downarrow\downarrow \downarrow\downarrow}\bra{\downarrow\downarrow\downarrow\downarrow} +\frac{2}{5} \ket{\psi_{\mathrm{d}}} \bra{\psi_{\mathrm{d}}}
\label{eq:steadystate}
\end{equation}
Therefore, for $N_B=2$, we have $x_{\mathrm{d}}=2/5$. 

We are interested in the entanglement between spins in $A$ and $C$ and from \eqref{eq:steadystate}, we can trace out the two middle spins belonging to domain $B$ and be left with the reduced denisty matrix of the spins in $A$ and $C$. This is given by:
\begin{equation} 
\rho_{AC}=\frac{17}{25} \ket{\downarrow\downarrow}\bra{\downarrow\downarrow} + \frac{8}{25} \ket{\Psi^+} \bra{\Psi^+}
\end{equation}
with $\ket{\Psi^+}=\frac{1}{\sqrt{2}} \left( \ket{\uparrow\downarrow}+\ket{\downarrow\uparrow}\right)$. In fact, the form of the reduced density matrix for this problem is always:
\begin{equation}
\rho_{AC}= x \ket{\Psi^+} \bra{\Psi^+} +\left(1-x\right) \ket{\downarrow\downarrow}\bra{\downarrow\downarrow}
\label{eq:rhoACss}
\end{equation}
 While the parameter $x_{\mathrm{d}}$ in \eqref{eq:steadyNb} describes the proportion of the state that remains in the dark state, the parameter $x$ in \eqref{eq:rhoACss} describes the proportion of the reduced state that is in the entangled Bell state, specifically, the triplet state. As shown above, for $N_B=2$, $x=8/25$. This analysis can be repeated for $N_B>2$ with the crucial step being defining the dark state \eqref{eq:darkNb} and the two other states $\ket{\psi_1}$ and $\ket{\psi_2}$ and then using such states to express the initial state.  While the formula for the dark state is given in \eqref{eq:darkNb}, the states $\ket{\psi_1}$ and $\ket{\psi_2}$ can be generalised for higher spin populations $N_B$ in a straightforward way. For $\ket{\psi_1}$, we have 
\begin{equation}
\begin{split}
&\ket{\psi_1}=  \frac{1}{\sqrt{2+4 N_B}} 
\\&\times  \Bigg\{
\ket{\uparrow}_A  \ket{\downarrow\cdots\downarrow}_B  \ket{\downarrow}_C
+
\ket{\downarrow}_A  \ket{\downarrow\cdots\downarrow}_B  \ket{\uparrow}_C
\\
&+2  \bigg[ \ket{\downarrow}_A 
 \underbrace{  \left( \ket{\uparrow \downarrow\cdots \downarrow}_B + \cdots +\ket{\downarrow\cdots \downarrow\uparrow}_B  \right) }_{\text{all permutations of a single spin up in }B } \ket{\downarrow}_C
\bigg]
\Bigg\}
\end{split}
\label{eq:psi1Nb}
\end{equation}

while for $\ket{\psi_2}$ we have:
\begin{equation}
\ket{\psi_2}=  \frac{1}{\sqrt{2} } \left(  \ket{\uparrow}_A \ket{\downarrow\cdots \downarrow}_B \ket{\downarrow}_C  - \ket{\downarrow}_A\ket{\downarrow\cdots \downarrow}_B\ket{\uparrow}_C  \right)
\end{equation}
With such definitions, one needs to express the initial state in terms of $\ket{\psi_{\mathrm{d}}}$, $\ket{\psi_1}$ and $\ket{\psi_2}$ and from this we know that the dark state component does not decay while the $\ket{\psi_1}$ and $\ket{\psi_2}$ components decay fully to the ground state.  Thus we can find the proportion of the state that remains in the dark state, and subsequently identify the dependence of $x_{\mathrm{d}}$ on $N_B$.  We shall summarise the findings here: The steady state of the total system is given by \eqref{eq:steadyNb}, with the dark state defined as in \eqref{eq:darkNb}, and with 
\begin{equation}
x_{\mathrm{d}}=\frac{N_B}{2 N_B+1}
\end{equation}
For the reduced system of two single spins in domains $A$ and $C$, the steady state is defined as in \eqref{eq:rhoACss} with 
\begin{equation}
x=\frac{2 N_B^2}{\left( 2 N_B+1\right)^2}
\end{equation}
\begin{figure}\captionsetup{position=top}
\centering
{\includegraphics[width=0.99\linewidth]{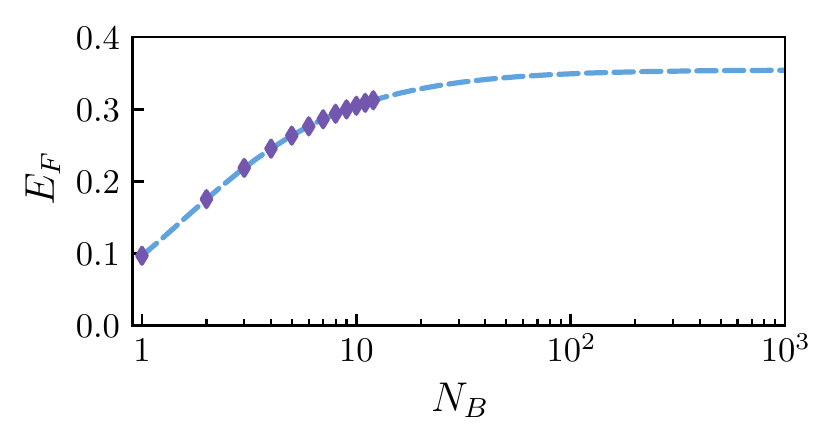}} 
\caption{\cjosephine{Steady state entanglement of formation between single spins in $A$ and $C$ for the initial condition \eqref{eq:alt-is} evolving under master equation \eqref{eq:masterABC} . The purple diamonds indicate data points acquired by numerically solving the master equation which can be done for $1\leq N_B \leq 12$, while the blue dashed line is \eqref{eq:eof} with the concurrence \eqref{eq:conc}}.}
\label{fig:eofaupNb}
\end{figure}
To quantify the entanglement of the two qubit reduced state, we again use the concurrence, which for this state \eqref{eq:rhoACss} is simply: $C=x$ and thus the concurrence of the two qubit state for a given $N_B$ is
\begin{equation}
C=\frac{2 N_B^2}{\left( 2 N_B+1\right)^2}
\label{eq:conc}
\end{equation}
and the entanglement of formation can be calculated via:
\begin{equation}
\begin{split}
\mathcal{E} = -&\left( \frac{1+\sqrt{1-C^2}}{2} \right)\log_2 \left(\frac{1+\sqrt{1-C^2}}{2} \right) 
\\
& -\left(\frac{1-\sqrt{1-C^2}}{2} \right) \log_2 \left(\frac{1-\sqrt{1-C^2}}{2} \right)
\end{split}
\label{eq:eof}
\end{equation}
In Fig.~\ref{fig:eofaupNb}, we show the behavoir  of the entanglement of formation for the range of $1\leq N_B\leq10^3$.  As the system size $N_B$ increases, so too does the proportion of the state that remains in the Bell state. For $N_B\to\infty$, this value goes to $x_\infty=1/2$. This means that the steady state of the two qubits in this limit is given by 
\begin{equation}
\rho_{AC}^\infty= \frac{1}{2} \ket{\Psi^+} \bra{\Psi^+} +\frac{1}{2} \ket{\downarrow\downarrow}\bra{\downarrow\downarrow} ,
\end{equation}
and the entanglement of formation for this state is 
\begin{equation}
\mathcal{E}_\infty \approx 0.354 \, .
\end{equation}
}

\end{document}